\documentclass[aps,reprint,showpacs,floatfix,amsmath,amssymb,pre]{revtex4-1}

 \usepackage{color} 
\usepackage{graphicx}

\usepackage{lmodern}						

\begin{document}

\title{Approximate solution to the stochastic Kuramoto model}
\author{Bernard Sonnenschein
}
\author{Lutz Schimansky-Geier}
\affiliation{Department of Physics, Humboldt-Universit\"at zu Berlin, Newtonstrasse 15, 12489 Berlin, Germany;}
\affiliation{Bernstein Center for Computational Neuroscience Berlin, Philippstrasse 13, 10115 Berlin, Germany;}

\begin{abstract}
We study Kuramoto phase oscillators with temporal fluctuations in the frequencies.
The infinite-dimensional system can be reduced in a Gaussian approximation to two first-order differential equations.
This yields a solution for the \emph{time-dependent} order parameter, which characterizes the
synchronization between the oscillators. The known critical coupling strength is exactly 
recovered by the Gaussian theory. Extensive numerical experiments further show that the analytical results are very accurate
below and sufficiently above the critical value. We obtain the asymptotic order parameter \emph{in closed form}, which suggests 
a tighter upper bound for the corresponding scaling. As a last point, we elaborate the Gaussian approximation in 
complex networks with distributed degrees.
\end{abstract}
\maketitle

\section{Introduction}
An often studied model describing the phenomenon of collective synchronization \cite{PikRosKu03,*BalJaPoSos10,AniAstNeVaSchi07} is due to Kuramoto \cite{Kur84}. 
It describes how the phases of coupled oscillators evolve in time. Being applicable to any system of nearly identical, weakly coupled 
limit-cycle oscillators, the Kuramoto model is way more than a toy model (for reviews see \cite{Str00,*AcBoViRiSp05}).
It is concerned with the competition between diversity, which hinders synchronization, and couplings, by which the oscillators tend to synchronize.
A topic that has been explored for a long time and where still many open questions remain, is the low-dimensional behavior that evidently hides behind the 
$N$-dimensional Kuramoto model (where the system size $N$ typically goes to infinity) 
\cite{KurNish86,*StrMiMat92,*WatStr93,*WatStr94,*PikRos08,*PikRos11,*MaBarStrOttSoAnt09,*MaMiStr09,OttAnt08,*OttAnt09,GiacPakPell12,Mir12,OmelWo13}.
It is this field of research where our work aims to contribute.
Specifically, we consider all-to-all coupled oscillators, where the diversity purely comes from noise acting on the frequencies. The goal is to find an evolution equation
for the order parameter and to obtain the corresponding solution. The latter should, at least
approximately, reveal the level of synchronization for any point in time and for any coupling strength. This is achieved here under the assumption that the phases of the 
oscillators are Gaussian distributed at all times \cite{KurSchu95,ZaNeFeSch03}. Such a procedure has also been used, e.g., for coupled FitzHugh-Nagumo oscillators 
\cite{TaPak01,*ZaSaLSGNei05,*HerTou12}, integrate-and-fire neurons \cite{Bu01} and networks of active rotators \cite{SonnZaNeiLSG13}.
After having obtained an expression for the order parameter, we examine its long-time asymptotic behavior. We finally present the extension to networks with distributed degrees.

\section{Model}
Consider a stochastic version of the Kuramoto model:
\begin{equation}
\dot{\phi}_i(t)=\xi_i(t)+\frac{K}{N}\sum_{j=1}^{N}\sin\left(\phi_j-\phi_i\right).
\label{model}
\end{equation}
The oscillators are indexed by $i=1,\ldots,N$ and $K$ stands for the coupling strength. All oscillators have the same constant natural 
frequency concerning the underlying limit-cycle. By virtue of the rotational symmetry 
in the model, we can subtract the natural frequency from the instantaneous 
frequencies without changing the dynamics. In this co-rotating frame the phases $\phi_i(t)$ describe the deviations from the
limit-cycle. 
Diversity among the oscillators is due to
stochastic forces $\xi_i(t)$ perturbing the evolution of the phases. Such time-dependent disorder is often modeled by Gaussian 
white noise \cite{LiGarNeiSchi04,AniAstNeVaSchi07}, which we also consider here. Therefore we have
\begin{equation}
 \begin{aligned}
  \langle\xi_i(t)\rangle&=0,\\
\langle\xi_i(t)\xi_j(t')\rangle&=2D\delta_{ij}\delta(t-t'),
 \end{aligned}
\end{equation}
where the angular brackets denote averages over different
realizations of the noise and the single nonnegative parameter $D$ scales its intensity. 
The noise terms $\xi_i(t)$ can be regarded as an accumulation of various stochastic processes,
such as the variability in the release of neurotransmitters or the quasi-random synaptic inputs from other neurons.
The case of dichotomous Markovian noise was studied in \cite{KoLuLSG02}.
One can set $D$ or $K$ to unity (by rescaling time), 
but for illustrative purposes we do not rescale \eqref{model}, a priori.
 
For non-identical oscillators without noise, the time-dependent functions $\xi_i(t)$ are replaced by time-independent natural frequencies $\omega_i$ that are drawn from
some frequency distribution $g(\omega)$. Such disorder is often called ``quenched''. The Kuramoto model with quenched disorder can be treated most elegantly by virtue of 
the Ott-Antonsen theory \cite{OttAnt08,*OttAnt09}. Remarkably, the latter provides a drastic but exact dimensionality reduction. A counterpart of the Ott-Antonsen theory 
needs to be found for the stochastic problem. We will exemplify an approximate method in order to get a low-dimensional dynamics.

\section{Theory}
In the following, we investigate the thermodynamic limit $N\rightarrow\infty$,
where the system is conveniently described
 by a probability density $\rho(\phi,t)$, which is normalized according
 to $\int_0^{2\pi}\rho(\phi,t)\mathrm d\phi=1\ \forall\ t$; 
 $\rho(\phi,t)$ $\mathrm d\phi$ gives the fraction of oscillators
 having a phase between $\phi$ and $\phi+\mathrm d\phi$ at time $t$.
 The completely asynchronous state is given by
 $\rho(\phi,t)=1/(2\pi)\ \forall\ t$.
 
 We start with the well-known nonlinear Fokker-Planck equation which governs the evolution of the one-oscillator probability
 density $\rho(\phi,t)$ \cite{Sak88}:
  \begin{equation}
   \frac{\partial\rho}{\partial t}\,=\,D\frac{\partial^2\rho}{\partial\phi^2}-\frac{\partial}{\partial\phi}\left[Kr\sin\left(\Theta-\phi\right)\, \rho\right]\ .
 \label{fpe}
 \end{equation}
 The mean-field amplitude $r(t)$ and phase $\Theta(t)$ involve $\rho(\phi,t)$, making the latter equation nonlinear in $\rho$:
 \begin{equation}
   r(t)\mathrm e^{i\Theta(t)}=\int_{0}^{2\pi}\mathrm d\phi'\ \mathrm e^{i\phi'}\ \rho\left(\phi',t\right)\, .
   \label{order_fpe}
 \end{equation}
 Since $\rho(\phi,t)$ is $2\pi$-periodic in $\phi$, we can write a Fourier series expansion
 \begin{equation}
 \rho(\phi,t)=\frac{1}{2\pi}\sum_{n=-\infty}^{+\infty}\rho_n(t)\mathrm e^{-in\phi}\ .
 \label{expansion} 
 \end{equation}
 Recall that $\rho(\phi,t)$ is a probability density, i.e. it is a normalized real quantity. Thus Eq. \eqref{expansion} is constrained by
$\rho_0=1$ and $\rho_{-n}=\rho_{n}^*$. Through the inverse transform of \eqref{expansion} one can also write
 \begin{equation}
 \rho_n(t)=\int_{0}^{2\pi}\mathrm d\phi'\ \rho\left(\phi',t\right)\ \mathrm e^{in\phi'}\equiv c_n(t)+is_n(t).
 \label{orderp}
 \end{equation}
Apparently, $n=1$ leads to Eq. \eqref{order_fpe}; in particular, the classical Kuramoto order parameter equals $r(t)=\left|\rho_1(t)\right|$.

\begin{figure}
\centering
\includegraphics[width=0.85\linewidth]{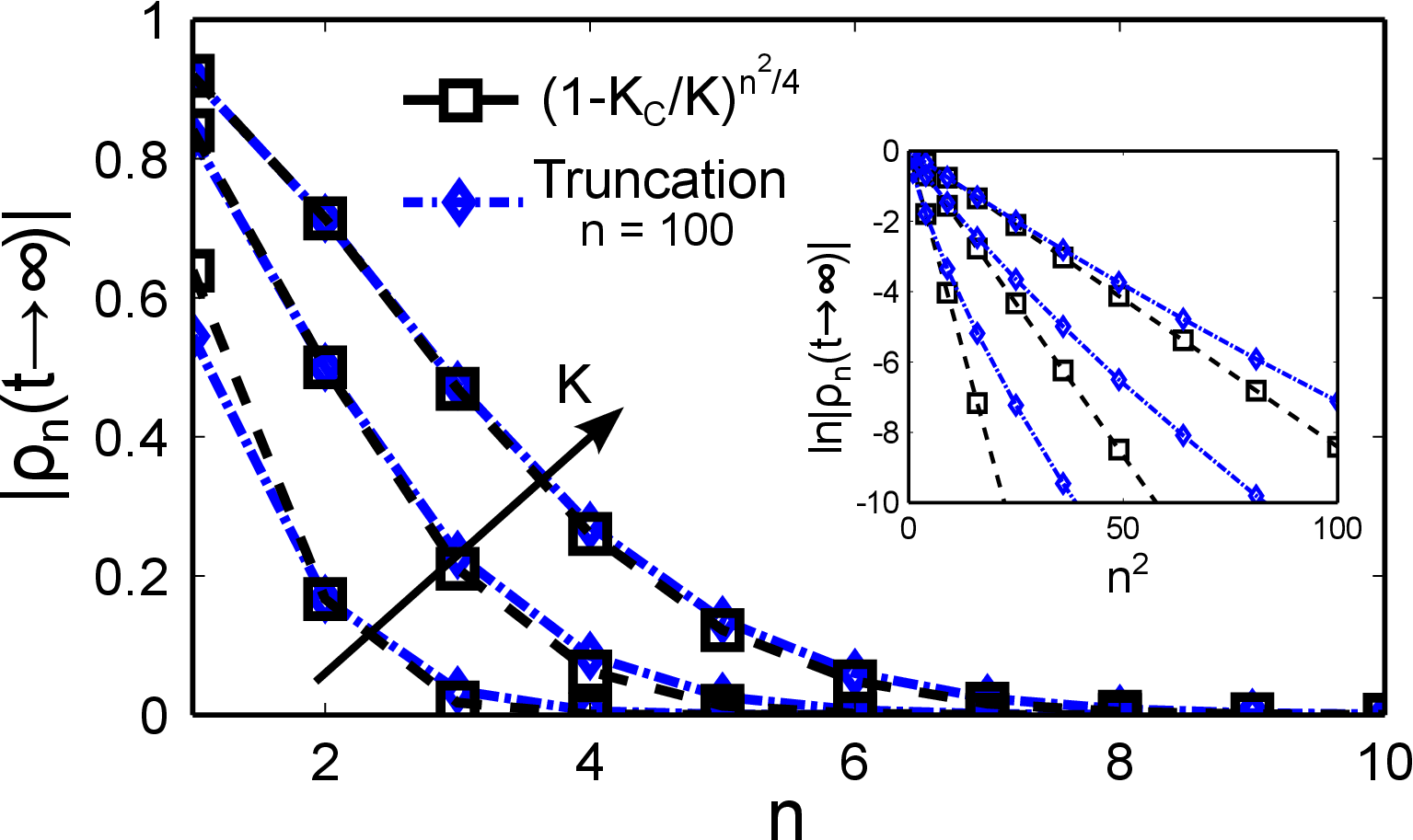}
\caption[]{(Color online) Decay of stationary Fourier amplitudes with order $n$. For clarity, points are connected by lines. 
Gaussian theory is compared with the numerical truncation of Eq. \eqref{chain} at $n=100$. Stationary values are 
taken at $t=500$ ($D=0.4$). Coupling strengths are $K=\left\{1.2K_c,2K_c,3.5K_c\right\}$. The inset shows the same data with
different axes.}
\label{rhon}
\end{figure}

Inserting \eqref{expansion} into \eqref{fpe} yields an infinite chain of coupled complex-valued equations for
 the Fourier coefficients \cite{SakShiKur88}
 \begin{equation}
 \frac{\dot{\rho_n}}{n}=\frac{K}{2}\left(\rho_{n-1}\rho_1-\rho_{n+1}\rho_{-1}\right)-Dn\rho_{n},
 \label{chain}
 \end{equation}
 with $n=1,2,\ldots,\infty$. 
 The crucial point now is that the coefficients $\rho_n(t)$ rapidly decay with increasing $n$, see Fig. \ref{rhon}.
 So one can easily obtain an approximate description of the underlying dynamics by truncating Eqs. \eqref{chain} at a large enough $n$. Typically,
 $n=6$ leads already to satisfactorily accurate results. We will use a much larger value when comparing theoretical with numerical solutions.
 
\section{Closure scheme}
\begin{figure}
 \centering
 \includegraphics[width=0.98\linewidth]{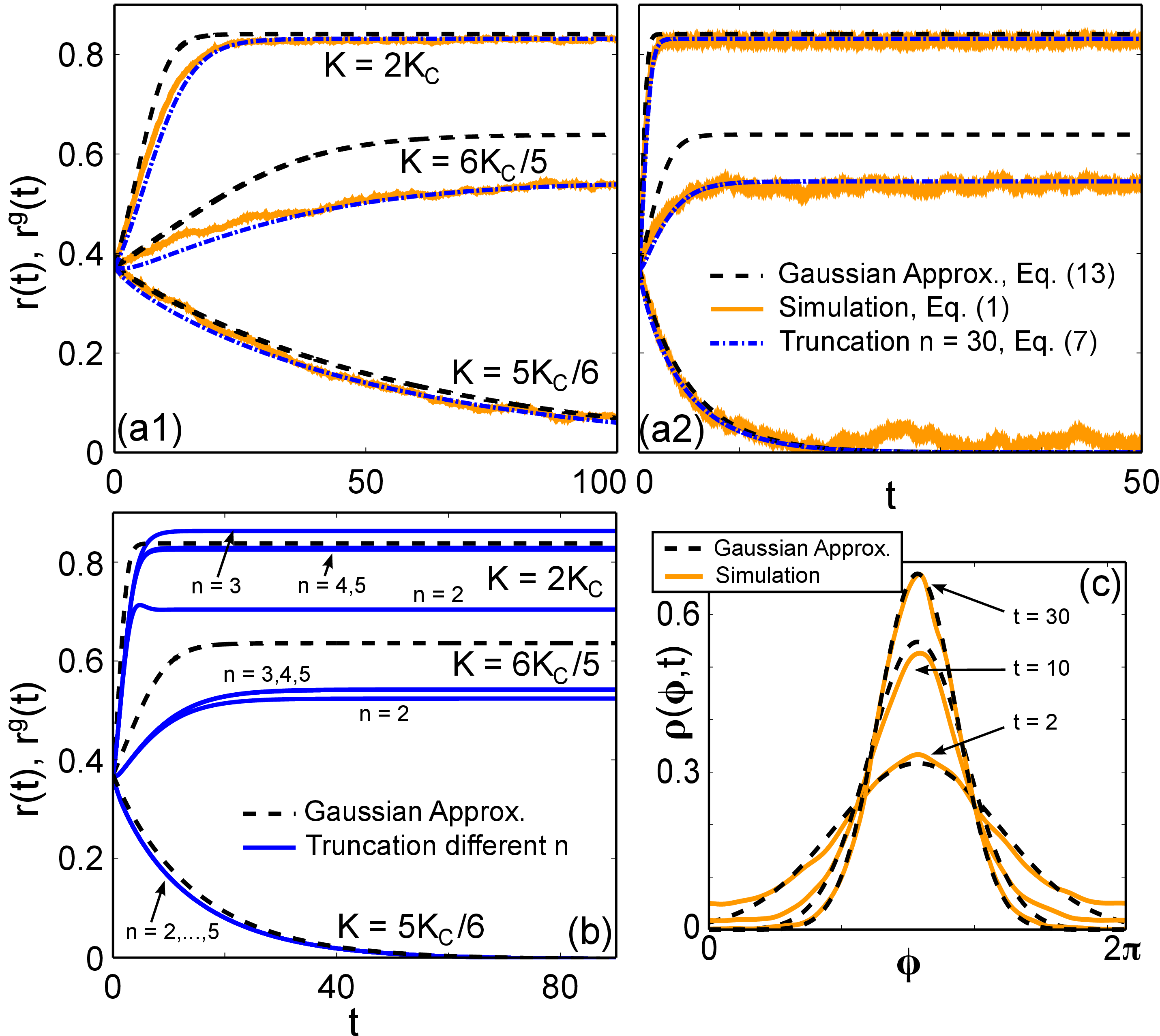}
 \caption[]{(Color online) Time evolution of Kuramoto order parameter with initial condition $r(0)=r^g(0)=0.373$. 
 (a1) compares theory and numerical experiments for different coupling strengths $K$ ($D=0.1$).
 The black dashed line shows the Gaussian theory $r^g(t)$, the blue dash-dotted line the numerical integration of truncated Eqs. \eqref{chain} at $n=30$, and
 orange solid line the simulation of full dynamics \eqref{model} with initially Gaussian distributed phases. In (a2) a different noise intensity is used, $D=1.2$.
 Simulation of full dynamics is performed with system size $N=15000$ and time steps $0.05$ for 
 $D=0.1$ and $0.005$ for $D=1.2$, respectively. (b) compares the accuracy between theory (dashed lines) and truncation 
 of \eqref{chain} at some $n$ (solid lines). (c) shows snapshots of phase distributions. For the theoretical lines the variance from Eq. \eqref{var} is used. 
 The initial variance is $\sigma^2(0)=1.96$.
 For all snapshots, phases are shifted by the same amount so that mean values are located at $\phi=\pi$. The parameters are the same as in (a1)'s uppermost curve.}
 \label{orderp_time}
 \end{figure}

Here we focus on a framework which is called the \emph{Gaussian approximation} \cite{ZaNeFeSch03}. There one assumes that the phases of the 
oscillators are Gaussian distributed
 with mean $m(t)$ and variance $\sigma^2(t)$ that are allowed to be time dependent. This approximates the bell-shaped curve found in numerical simulations 
 [Fig. \ref{orderp_time}(c)]. Then the real and imaginary parts of $\rho_n(t)$ turn into
 \begin{equation}
 \begin{aligned}
 c_n^{g}(t)&=\mathrm e^{-n^2\sigma^2(t)/2}\cos\left[nm(t)\right],\\
 s_n^{g}(t)&=\mathrm e^{-n^2\sigma^2(t)/2}\sin\left[nm(t)\right].
 \end{aligned}
 \label{cs}
 \end{equation}
 The superscript $g$ shall label expressions obtained within the Gaussian theory. Eqs. \eqref{cs} imply that the Fourier amplitudes are not affected by the mean phase:
\begin{equation}
 \left|\rho_n^{g}(t)\right|=\mathrm e^{-n^2\sigma^2(t)/2}.
\label{gaussian_rho}
\end{equation}
Interestingly, one can realize a similarity between the Gaussian ansatz for the Kuramoto model with noise
and the Ott-Antonsen ansatz for the case with quenched disorder instead of noise.
The latter namely consists of identifying the $n$th Fourier coefficient with the $n$th power of a specific complex function. 
It has to be emphasized that in contrast to the Gaussian ansatz, the Ott-Antonsen ansatz is exact (for limitations see \cite{Mir12} and for a 
recent generalization via a standing wave ansatz see \cite{OmelWo13}). 

 One can show by transformation of variables $\left\{c_1^g,s_1^g\right\}\rightarrow\left\{m,\sigma^2\right\}$ that the cumulants obey the differential equations
 \begin{equation}
 \dot{\sigma^2}=2D+K\left(\mathrm e^{-2\sigma^2}-1\right)
 \label{vardot}
 \end{equation}
 and $\dot m=0$ (compare with Refs. \cite{ZaNeFeSch03,SonnZaNeiLSG13}). So the mean of the phase distribution is in fact constant in time. This comes from the 
 rotational symmetry in the model, see above the discussion of Eq. \eqref{model}.

\section{Time-dependent solutions and long-time limits}
The differential equation \eqref{vardot} can be directly integrated after separation of variables and
the variance of the phases then reads
 \begin{equation}
 \begin{aligned}
 \sigma^2(t)=&\frac{1}{2}\ln\left[\frac{K}{K-2D}+\mathrm e^{-2t(K-2D)}\times\right.\\
 &\left.\times\left(\mathrm e^{2\sigma^2(0)}-\frac{K}{K-2D}\right)\right]
  \label{var}
  \end{aligned}
 \end{equation}
for $K\neq 2D$. Considering the long-time asymptotic limit $t\rightarrow\infty$, we see that the variance grows to infinity for $K<2D$, while it asymptotes to 
\begin{equation}
\sigma^2(t\rightarrow\infty)=\frac{1}{2}\ln\left(\frac{K}{K-2D}\right)
\label{sigma_end}
\end{equation}
for $K>2D$.
Hence, the variance equals zero, either if the coupling strength goes to infinity, $K\rightarrow\infty$, or if the noise intensity $D$ vanishes; $\sigma^2=0$ would correspond to
a perfectly synchronized state, while $\sigma^2>0$ signals partial synchronization, and the completely asynchronous state is characterized by a diverging variance. This means 
that at $K_c=2D$ the population of oscillators transitions from incoherent to partially synchronized behavior. 
Noteworthy, this critical value is \emph{exact}, as known for a long time \cite{Sak88}.
Now, by inserting \eqref{var} into \eqref{gaussian_rho}, we readily get an expression for all Fourier amplitudes:
 \begin{equation}
  \left|\rho_n^{g}(t)\right|=\left|\frac{K-K_c}{\left(\frac{K-K_c}{\left|\rho_n^{g}(0)\right|^{4/n^2}}-K\right)\mathrm e^{-2t(K-K_c)}+K}\right|^{n^2/4}.
 \label{all_order}
 \end{equation}
It is remarkable that we can report this approximate solution even though the effects of noise on the collective dynamics of phase oscillators were comprehensively studied
before \cite{Sak88,StrMir91,*Cra94,*ArPer94,*Craw95,*CraDa99,*PikRu99,*BalSa00,*TesScToCo07,BerGiaPak10,GiacPakPell12,SonnSchi12}. 

Equation \eqref{all_order} indicates that $\left|\rho_n^g(t)\right|$ 
 goes to zero exponentially fast with increasing time for subcritical coupling, while it goes to
 \begin{equation}
\left|\rho_n^{g}(t\rightarrow\infty)\right|=\left[1-(K_c/K)\right]^{n^2/4}
 \label{theory}
 \end{equation}
 for $K>K_c$. In Fig. \ref{rhon} we compare this analytical result with the numerical truncation of \eqref{chain}. We see that the decay of the Fourier coefficients with the order $n$
is captured by the Gaussian theory. Plotting 
$\ln\left|\rho_n(t\rightarrow\infty)\right|$ as a function of $n^2$ should give straight lines. The inset of
Fig. \ref{rhon} shows that the Gaussian approximation tends to underestimate the higher Fourier amplitudes.
The deviation in the vicinity of the critical coupling strength, visible for the lowest curve in the main part of Fig. \ref{rhon}, will be discussed in detail below. 

In what follows, we explore the classical Kuramoto order parameter, that is $r(t)=\left|\rho_1(t)\right|$ [see Eq. \eqref{orderp}]. It is illustrative
to derive the evolution equation for $r(t)$ directly in a slightly different way. The power of the Gaussian ansatz lies essentially in the fact that all $c_n^g$ and $s_n^g$ are given 
by $c_1^g$ and $s_1^g$: $c_2^g=(c_1^g)^4-(s_1^g)^4$, $s_2^g=2s_1^gc_1^g\left[(s_1^g)^2+(c_1^g)^2\right]$, etc. \cite{ZaNeFeSch03}. Thereby a closure in \eqref{chain} is achieved and one is left with
 \begin{equation}
 \begin{aligned}
   \dot c_1^{g}&=-Dc_1^{g}+\frac{Kc_1^{g}}{2}\left\{1-\left[\left(c_1^{g}\right)^2+\left(s_1^{g}\right)^2\right]^2\right\},\\
   \dot s_1^{g}&=-Ds_1^{g}+\frac{Ks_1^{g}}{2}\left\{1-\left[\left(c_1^{g}\right)^2+\left(s_1^{g}\right)^2\right]^2\right\}.
 \end{aligned}
 \label{dotcs}
 \end{equation}
 Now with $c_1^g=r^g\cos\theta^g$ and $s_1^g=r^g\sin\theta^g$ one obtains
 \begin{equation}
   \dot r^{g}=\frac{r^{g}}{2}\left\{K\left[1-\left(r^{g}\right)^4\right]-2D\right\}
 \label{r2}
 \end{equation}
 and $\dot\theta^{g}=0$. The solution of Eq. \eqref{r2} coincides, of course, with the $n=1$ solution \eqref{all_order}.

 In figure \ref{orderp_time}, the time-dependent order parameter is depicted [solution \eqref{all_order} with $n=1$]. While 
 panels (a1) and (a2) show how the analytical result compares with numerical experiments, panel (b) compares the Gaussian 
 approximation with truncation of \eqref{chain} at some $n$. We observe that the theory overestimates the order parameter,
 in particular in the non-stationary regime and slightly above the critical coupling $K_c$. This comes from the fact that the phases
 are more heavy tailed than predicted by the Gaussian theory, panel (c). Below $K_c$, the theoretical 
 lines are close to the numerically generated ones. Furthermore, for sufficiently strong
 couplings, the theory clearly outperforms numerical truncation. This can be seen in panel (b) for 
 $K=2K_c$. Note that $n=3$ corresponds already to six coupled differential equations, while the Gaussian ansatz leads
 to a two-dimensional system.

\section{Discussion of scaling}
We proceed with discussing the long-time behavior of the order parameter, i.e. Eq. \eqref{theory} with $n=1$.
Let $r_0$ denote the stationary value from now on.
In figure \ref{rInf} panel (a) we compare our scaling result with numerical experiments. In agreement with our observations concerning the time-dependent order parameter
(Fig. \ref{orderp_time}), there is a deviation for coupling constants in the interval $K_c<K\lesssim2K_c$. This is due to the fact that Eq. \eqref{theory} violates the 
square-root scaling law. The $(K-K_c)^{1/2}$ scaling is well-known \cite{Sak88}; for completeness, we show the crossover to the square-root scaling
in the inset of panel (a). While this is not captured by Eq. \eqref{theory}, the analytical scaling result is highly accurate for 
sufficiently large coupling strengths, i.e. $K\gtrsim 2K_c$. 

We remark that the long-time asymptotic order parameter $r_0$ is exactly
 given by a transcendental equation,
 \begin{equation}
 r_0=\frac{I_1(r_0K/D)}{I_0(r_0K/D)},
\label{trans1}
 \end{equation}
where $I_0$ and $I_1$ denote modified Bessel functions of the first kind of order $0$ and $1$, respectively \cite{BerGiaPak10}.
 Recently, besides other interesting results, Bertini \emph{et al.}
 proved certain bounds for the asymptotic order parameter, namely $\left(1-K_c/K\right)^{1/2}<r_0<\left(1-K_c/2K\right)^{1/2}$ 
 \cite{BerGiaPak10}. Now, with the general form of Bernoulli's inequality, we find $\left(1-K_c/K\right)^{1/4}<\left(1-K_c/2K\right)^{1/2}$.
 Indeed, $(1-K_c/K)^{1/4}$ seems to establish an improved upper bound. This is shown in Fig. \ref{rInf} panel (b), where we depict the bounds proven 
 by Bertini \emph{et al.} in comparison with our scaling result and the ``exact'' solution, obtained from Eq. 
 \eqref{trans1} via numerical determination of the roots. In the inset one can appreciate the accuracy of our result.

We would like to mention the recent work \cite{Cha13}, where, based on our derived expressions above and previous results \cite{ChaVattBou05}, an unifying fitting procedure is proposed,
such that the square-root scaling law is recovered.

\section{Temporal fluctuations vs. quenched disorder}
Kuramoto showed for a general symmetric and unimodal frequency distribution $g(\omega)$ that $r_0$ satisfies \cite{Kur84}
\begin{equation}
1=K\int_{-\pi/2}^{\pi/2}\cos(\phi)^2 g\left[r_0 K\sin(\phi)\right]\mathrm d\phi\ .
\label{self}
\end{equation}
From expanding $g\left[r_0K\sin(\phi)\right]$ in powers of $r_0K$, he then found a square-root scaling law for $r_0$ as $K\rightarrow K_c$. Notably, Kuramoto also
 \begin{figure}
 \centering
 \includegraphics[width=0.76\linewidth]{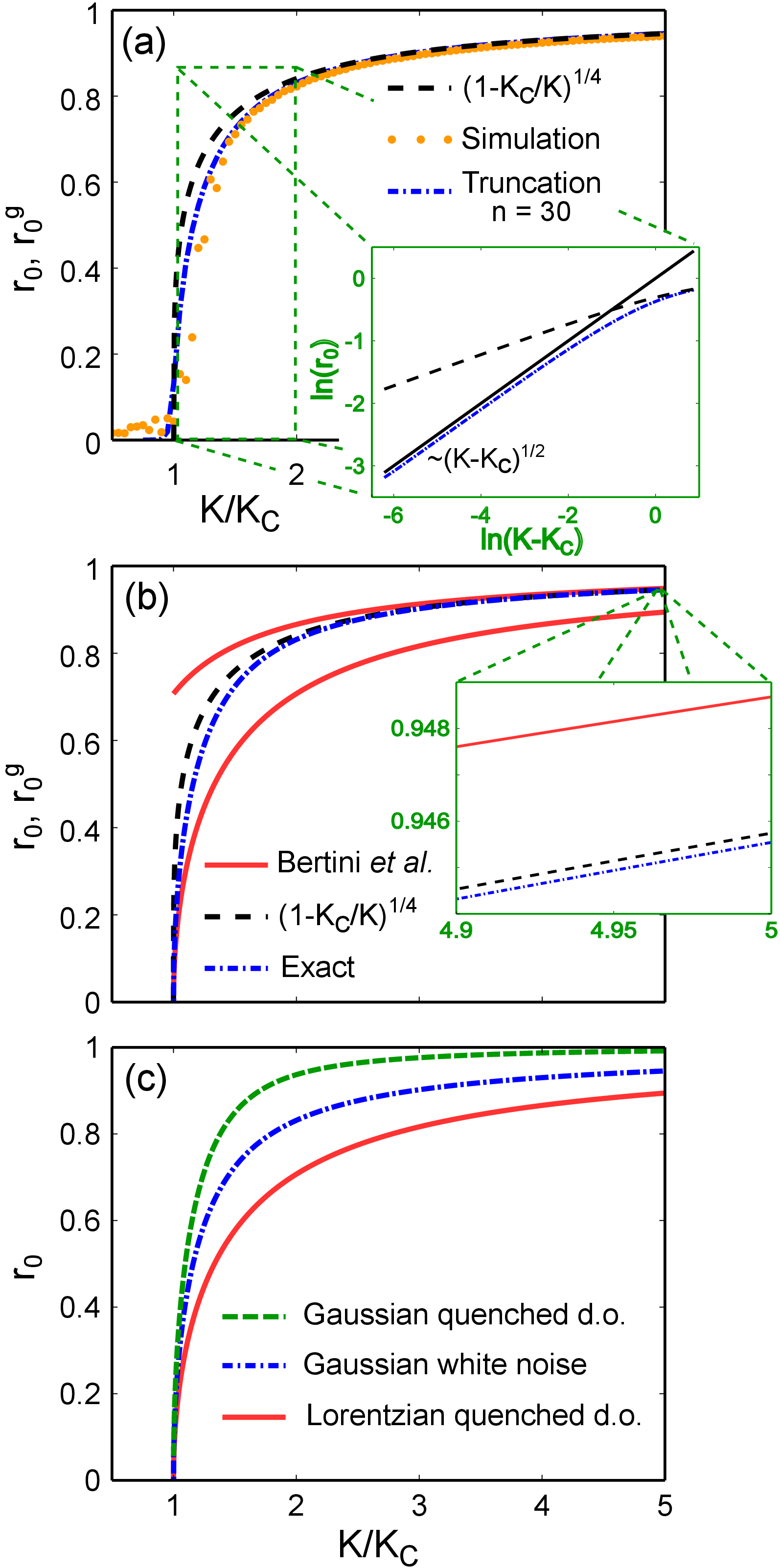}
 \caption[]{(Color online) Scaling of the asymptotic order parameter as a function of coupling strength divided by critical value, $K/K_c$. In (a) theory \eqref{theory} for $n=1$, $r_0^g$,
 is compared with numerical experiments [numerical integration of \eqref{model} with $N=5000$, $D=0.1$, integration step $0.05$ and time-averaged between $t=100$ and $t=200$; 
 truncated system \eqref{chain} is integrated until $t=1500$ and the final value for the order parameter is taken]. The inset shows square-root scaling slightly above critical coupling, which is not
 covered by the theory; the smallest value there $K-K_c=0.002$.
 In (b) theory is compared with the exact solution, Eq. \eqref{trans1}, and with results from \cite{BerGiaPak10}. The inset shows a zoom-in. (c) compares levels
 of synchronization for various sources of disorder.}
 \label{rInf}
 \end{figure}
found that in case of a Lorentzian $g(\omega)$, the order parameter equals exactly $r_0=\sqrt{1-K_c/K}$ for all $K>K_c$. Note that for a Gaussian $g(\omega)$ with standard 
deviation $\sigma$, \eqref{self} can be brought into
a form that is similar to \eqref{trans1}, namely a transcendental equation with the aforementioned Bessel functions:
\begin{equation}
\exp\left(\frac{r_0^2 K^2}{4\sigma^2}\right)\sqrt{\frac{8\sigma^2}{\pi K^2}}=I_0\left(\frac{r_0^2K^2}{4\sigma^2}\right)+I_1\left(\frac{r_0^2K^2}{4\sigma^2}\right)
\label{trans2}
\end{equation}
The comparison between Gaussian and Lorentzian quenched disorder and Gaussian white noise is given in Fig. \ref{rInf} panel (c). With respect to $K/K_c$, 
the Lorentzian quenched disorder hinders synchronization most strongly, then in the middle comes Gaussian white noise, and Gaussian quenched disorder 
hinders synchronization in the weakest form.

\section{Complex networks}
Finally, we discuss the extension to complex coupling structures, which makes the problem intractable in general.
The remedy lies in finding a suitable approximative description. Let $A$ be the adjacency matrix: $A_{ij}=1$ 
if node $j$ couples to node $i$ and $A_{ij}=0$ if this is not the case. Let further $k_i$ denote the degree, the number of connections of node $i$. Then for an undirected
network where the adjacency matrix is symmetric, one can approximate the latter by $\tilde A_{ij}=k_i k_j/\sum_l k_l$. By this coarse-graining one can satisfactorily 
tackle the problem in mean-field approximation \cite{Ichi04,*ResOttHu05,SonnSchi12}. As a result, all nodes with the same degree $k$ build a subpopulation with their own mean-field
variables $r_k(t),\Theta_k(t)$: 
\begin{equation}
  r_k(t)\mathrm e^{i\Theta_k(t)}=\int_{0}^{2\pi}\mathrm d\phi'\ \mathrm
  e^{i\phi'}\ \mathcal \rho\left(\phi',t|k\right).
  \label{orderp_k}
\end{equation}
The difference to the all-to-all uniformly coupled case [Eq. \eqref{order_fpe}] appears in the one-oscillator probability density $\rho\left(\phi,t|k\right)$, which now depends on the individual degree $k$. The subpopulations are coupled through averages over the degree distribution $P(k)$, such that the global mean-field variables are given by
\begin{equation}
 r(t)\mathrm e^{i\Theta(t)}=\langle k'r_{k'}(t)\mathrm e^{i\Theta_{k'}(t)}\rangle/\langle k'\rangle.
 \label{rkthetak}
\end{equation}
Here we use the notation $\langle\ldots\rangle\equiv\sum_{k'}\ldots P(k')$. In Gaussian approximation
we find (compare with Ref. \cite{SonnZaNeiLSG13})
\begin{equation}
 \begin{aligned}
\dot r_k^g&=\frac{\left(1-r_k^g\right)^4}{2N\langle k'\rangle}\ K\ k\  \langle k' r_{k'}^g\cos\left(\Theta_{k'}^g-\Theta_k^g\right)\rangle-r_k^g D, \\
\dot \Theta_k^g&=\frac{\left(r_k^g\right)^{-1}+\left(r_k^g\right)^3}{2N\langle k'\rangle}\ K\ k\ \langle k' r_{k'}^g\sin\left(\Theta_{k'}^g-\Theta_k^g\right)\rangle.
 \end{aligned}
 \label{dotorder}
\end{equation}
We may set $\Theta_k^g(t)=0$ in the co-rotating frame (all oscillators have zero average frequency).
Then the order parameter becomes $r^g=\langle k' r_{k'}^g\rangle/\langle k'\rangle$.
Instead of Eq. \eqref{r2} one obtains
\begin{equation}
    \dot r^{g}=\frac{K}{2N\langle k'\rangle}\left\langle k'^2\left[1-\left(r_{k'}^g\right)^4\right]\right\rangle r^g-Dr^g.
 \label{r_networks}
\end{equation}
Near the synchronization transition the contributions of $\left(r_{k}^g\right)^4$ can be neglected. So at $K_c=2DN\langle k'\rangle/\langle k'^2\rangle$ the 
order parameter switches from exponentially decreasing to increasing. Again, the Gaussian approximation reproduces the known critical coupling strength, see 
Ref. \cite{Ichi04,*ResOttHu05} without and Ref. \cite{SonnSchi12} with additive noise.

\section{Conclusion}
We have investigated the stochastic Kuramoto model, where the only source of disorder comes from temporal fluctuations acting on the evolution 
of the oscillator phases. In the continuum limit the system is infinite dimensional. By assuming Gaussianity in the phase distribution at all times, 
we found an approximate reduced description, which consists of two uncoupled ordinary first-order differential equations. As a consequence, we could easily find a full
solution. Specifically, we derived an expression for the time-dependent order parameter, which reveals the level of synchronization for any point in time and for any coupling strength. 
Noteworthy, the critical coupling strength for the onset of synchronization
is exactly reproduced by our theory. We also found that the Gaussian approximation is accurate for a coupling weaker or twice as strong as the critical one. 
In the vicinity of the critical value, the Gaussian theory does not reproduce the square-root scaling. Remarkably, however, the obtained scaling law appears to be an improved explicit upper bound
for the stationary order parameter. It is also interesting to see that the Gaussian approximation provides a simple way to calculate analytically the synchronization transition point in 
complex networks.
By showing where the Gaussian ansatz is valid, and where it is incorrect, our work sheds light on the underlying low-dimensional
dynamics. We believe that this is a promising direction for future research.

\acknowledgments
Thanks to O. E. Omel'chenko for stimulating conversations and P. K. Radtke for comments on the manuscript.
Work was supported by the Deutsche Forschungsgemeinschaft (GRK1589/1) and project A3 of the Bernstein Center for Computational Neuroscience Berlin.

\bibliography{bibliography}
\end{document}